\documentclass[twocolumn, a4paper, 11pt,book]{archive}
\usepackage{lineno,hyperref}
\modulolinenumbers[5]
\modulolinenumbers[1]

\usepackage{color}
\usepackage{multirow,bigdelim}
\usepackage{version}
\usepackage{graphicx}

\usepackage[]{natbib}

\usepackage{subcaption}
\usepackage{graphicx}
\usepackage{lscape}   
\usepackage{txfonts}
\usepackage{enumitem}   
\usepackage{xcolor}
\definecolor{darkorange}{rgb}{1.0, 0.55, 0.0}
\definecolor{britishracinggreen}{rgb}{0.0, 0.26, 0.15}
\definecolor{bondiblue}{rgb}{0.0, 0.58, 0.71}

\definecolor{klein}{rgb}{0.54, 0.17, 0.89}

\usepackage{soul}

\usepackage{tikz}
\usetikzlibrary{positioning}
\usepackage{pgfplots}
\usepackage{amsmath,amssymb,amsfonts}
\usepgfplotslibrary{fillbetween}
\begin{document} 

\twocolumn[{%
 \centering

  {\center \bf \huge The Flying Saucer edge-on disc’s Near Infrared silhouette revealed by the JWST JEDIce program}\\
%   \subtitle{I. Overviewing the $\kappa$-mechanism}
\vspace*{0.25cm}

   {\Large Emmanuel Dartois \inst{1}, Jennifer A. Noble \inst{2}, Jennifer B. Bergner \inst{3}, Klaus M. Pontoppidan \inst{4}, Korash Assani  \inst{5}, Daniel Harsono \inst{6}, Melissa K. McClure \inst{7}, Julia C. Santos \inst{8}, Will E. Thompson \inst{3}, Lukas Welzel \inst{7}, Nicole Arulanantham \inst{9}, Alice S. Booth \inst{8}, Maria N. Drozdovskaya \inst{10}, Zhi-Yun Li \inst{5}, Jie Ma \inst{11}, Laurine Martinien \inst{11}, Fran\c{c}ois M\'enard \inst{11}, Karin Oberg \inst{8}, Karl Stapelfeldt \inst{4}, Yao-Lun Yang \inst{12}}
             \vspace*{0.25cm}
             
$^1$ Institut des Sciences Mol\'{e}culaires d’Orsay, CNRS, Univ. Paris-Saclay, 91405 Orsay, France\\
\email{emmanuel.dartois@universite-paris-saclay.fr}\\
$^2$ Physique des Interactions Ioniques et Mol\'{e}culaires, CNRS, Aix Marseille Universit\'e, Marseille, France\\
$^3$ Dept. Chemistry, University of California, Berkeley, CA, USA\\
$^4$ Jet Propulsion Laboratory, California Institute of Technology, 4800 Oak Grove Drive, Pasadena, CA 91109, USA\\
$^5$ Astronomy Department, University of Virginia, 530 McCormick Rd., Charlottesville, VA 22903\\
$^6$ Institute of Astronomy, Department of Physics, National Tsing Hua University, Hsinchu, Taiwan\\
$^7$  Leiden Observatory, Leiden University, Leiden, The Netherlands \\
$^8$ Center for Astrophysics, Harvard \& Smithsonian, 60 Garden St., Cambridge, MA 02138, USA \\
$^9$ Astrophysics \& Space Institute, Schmidt Sciences, New York, NY 10011, USA\\
$^{10}$ Physikalisch-Meteorologisches Observatorium Davos und Weltstrahlungszentrum, Dorfstrasse 33, 7260, Davos Dorf, Switzerland\\
$^{11}$ Univ. Grenoble Alpes, CNRS, IPAG, 38000, Grenoble, France\\
$^{12}$ Star and Planet Formation Laboratory 2-1 Hirosawa, Wako, Saitama 351-0198, Japan\\

\vspace*{0.5cm}
{December 15, 2025}\\

{\it \huge Accepted for publication in Astronomy \& Astrophysics}\\
 \vspace*{0.5cm}

}]

  \section*{Abstract}
   {Edge-on discs offer a unique opportunity to probe radial and vertical dust and gas distributions in the protoplanetary phase. This study aims to investigate the distribution of micron-sized dust particles in the Flying Saucer %(BKLT J162813-243139) 
   in Rho Ophiuchi,
   leveraging the unique observational conditions of a bright infrared background that enables the edge-on disc to be seen in both silhouette and scattered light at certain, specific wavelengths.} 
   {   As part of the JWST Edge-on Disc Ice program ('JEDIce'), we use NIRSpec IFU observations of the Flying Saucer, serendipitously observed against a PAH-emitting background, to constrain the dust distribution and grain sizes through radiative transfer modelling.
   }
    {Observation of the Flying Saucer in silhouette at 3.29 $\mu$m reveals that the midplane radial extent of small dust grains is $\sim$235 au, larger than the large-grain disc extent previously determined to be 190 au from millimetre data. The scattered light observed in emission probes micron sized icy grains at large vertical distances above the midplane. The vertical extent of the disc silhouette is similar at visible, near-IR, and mid-IR wavelengths, corroborating the conclusion that dust settling is inefficient for grains as large as tens of microns, vertically and radially.}\\

   {Keywords: Protoplanetary discs / radiative transfer / scattering  / planets and satellites: formation / dust, extinction}
%
%%%%%%%%%%%%%%%%%%%%%%%%%%%%%%%%%%%%%%%%%%%%%%%%%%%%%%%%%%%%%%
\section{Introduction}
Protoplanetary discs with an `edge-on' viewing geometry are oriented with the disc occulting the star along our line of sight, providing a unique view of the disc's vertical structure \citep[e.g.][]{Burrows1996,Watson2007,Lin2023,Duchene2024,Dartois2025,Ballering2025}. The so-called `Flying Saucer' edge-on disc (BKLT J162813-243139) was discovered at the periphery of the $\rho$ Ophiuchi cloud complex in the near-infrared (NIR; J, H, and Ks bands) by \cite{Grosso2001,Grosso2003}. In the millimetre, this edge-on disc appears in silhouette against the CO $\rm J=2-1$ emission from the background $\rho$ Oph molecular cloud complex \cite{Guilloteau2016}. From these NIR to mm studies, the inclination of the Flying Saucer is estimated to be $\rm i \gtrsim 86^o$. Using Spitzer mid-IR spectroscopic data and additional near-IR and mid-IR photometric constraints, \cite{Pontoppidan2007} found that including large grains up to ten micrometres in size in the disc surface layers was required to explain the scattered light level in the spectral energy distribution for this disc. In this letter, 
we focus on the newly revealed NIR silhouette of the Flying Saucer disc, which is -- at some wavelengths -- observed both in emission and extinction by JWST NIRSpec.
   
%%%%%%%%%%%%%%%%%%%%%%%%%%%%%%%%%%%%%%%%%%%%%%%%%%%%%%%%%%%%%%%
%
\section{Observations}
The observations presented here are part of the JWST Edge-on disc Ices program (JEDIce, GO \#5299, PI J. Bergner). 
%_____________________________________________________________
%                        A figure as large as the column width
%-------------------------------------------------------------
   \begin{figure}[h!]
   \centering
   \includegraphics[width=0.8\columnwidth, trim={0cm 0cm 0cm 0},clip]{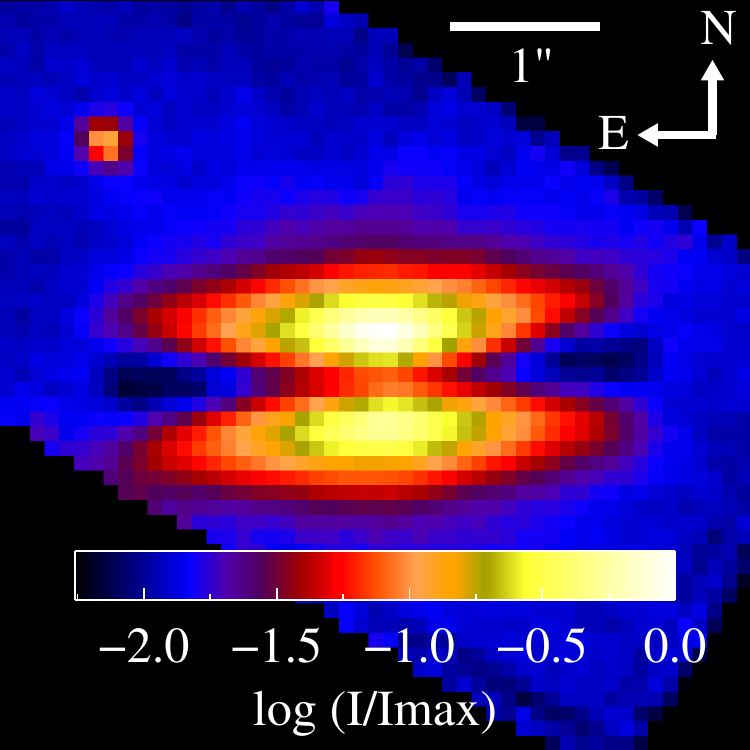}
      \caption{NIRSpec spectroscopic imaging data at 3.29$\pm$0.015~$\mu$m (i.e. stacking channels at the peak of the PAH band). The disc is observed both in emission (lobes, red-yellow) and in silhouette (midplane, black) against the ambient field emission arising from the CH stretching mode of UV-excited astro-PAHs (blue).}
         \label{fig1}
         \vspace*{-0.5cm}
   \end{figure}
%-------------------------------------------------------------
We summarise, in Sect.~2, the observations and the specific wavelengths where the disc appears in absorption against the background.  In Sect.~3, we present radiative transfer models of dust grain distributions and the structure of the disc in order to compare with the observations. We discuss the implications for the disc's dust sizes and settling in Sect.~4.
BKLT J162813-243139 (Flying Saucer) was observed with NIRSpec on August 25th, 2024. The observations were taken with the G235H and G395H filters from 1.66--3.17 $\mu$m and 2.87--5.27 $\mu$m, respectively, at a spectral resolution of R$\sim$2700, using the IFU observing mode with a pixel scale of 0.1$\arcsec$~pixel$^{-1}$ and a FOV of 3$\arcsec \times 3\arcsec$. A two-tile mosaic was used to capture the full extent of the disc.  For both the G235H and G395H observations, we used a small cycling dither pattern with 12 points and the NRSIRS2RAPID readout pattern.  
25 (10) groups per integration were used for the G395H (G235H) observations, for a total exposure time of 4552 (1926) sec.  
Data were reduced using calibration software version 1.17.1 and reference file database jwst\_1322$.$pmap.
The Flying Saucer disc, lying at the edge of the $\rho$ Oph cloud complex, is surrounded by an extended ambient emission field of infrared fluorescence \citep{Leger1984} from astro-PAH carriers exposed to ultraviolet photons. Within the NIRSpec cube, the disc is observed not only via the scattered light from its upper layers, but also in extinction close to the disc midplane, at the characteristic wavelengths corresponding to strong PAH emission features.  
This is particularly evident at around 3.29 $\mu$m, as shown in Fig.\ref{fig1}. A second object, most likely a field star, can be seen in the NE of the field and is not considered hereafter.
We complement the NIRSpec data with archive data spanning the visible to the millimetre. We retrieved archival data from HST (F475W, 0.47$\mu$m, program \#13643), JWST MIRI/Imager (F770W, 7.7$\mu$m, program \#4290), and JWST MIRI/MRS (Channel 3, program \#1280), in which the disc is also seen in extinction against a background of astro-PAH or H$_2$ emission.  We also include ALMA Band 7 continuum data (cycle 9 program \#2022.1.00742.S), in which the disc is observed in emission.  For the JWST MIRI/MRS datacube, three imaging frames were stacked around the peak of the H$_2$ line at 17.035$\mu$m to reproduce a narrow band imaging filter. 
%_____________________________________________________________
%                        A figure as large as the column width
%-------------------------------------------------------------
\begin{figure*}[ht!]
\centering
\includegraphics[width=2.05\columnwidth, trim={0 0cm 0cm 10cm},clip]{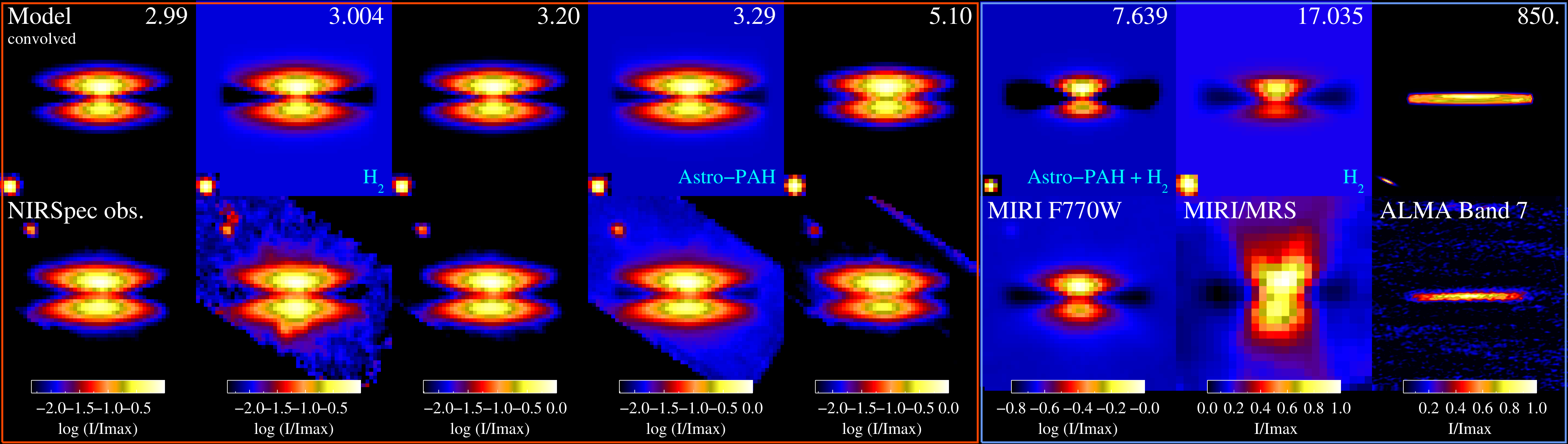}
\caption{Benchmark model of the Flying Saucer from near-IR to mm, compared to observations on a $\rm 5^{\prime\prime} \times5^{\prime\prime}$ field of view. The upper row shows the model images, convolved to observed spatial resolutions, and the bottom row displays the observations. Selected wavelengths across the NIRSpec range are shown along with archival images at 7.639$\mu$m (MIRI/Imager F770W filter pivot wavelength), 17.035$\mu$m (MIRI/MRS) and millimetre (ALMA) wavelengths. 
The contribution of the H$_2$ disc wind at 3.004 and 17.035$\mu$m was not included in the model.
The components dominating the ambient emission field at each wavelength are labelled astro-PAH and H$_2$ on the corresponding images. Images were rotated by 3$^\circ$ to account for the position angle.}
\label{fig_modele}
\end{figure*}
%-------------------------------------------------------------
These archive images are shown in Fig.\ref{fig_archive_data}. The 17.035$\mu$m synthetic narrow band filter provides an important constraint because it is the longest wavelength with little or no dust settling at which the disc silhouette is observed, as will be shown below. 
The HST and JWST wideband imaging observations provide additional information, limited by the fact that the signal from the various contributing components is integrated across the whole filter bandwidth.
We use the broadband images at 7.7$\mu$m and in the millimetre to show the continuity in the MIR and compare to the observed dust settling, respectively.  The focus of our analysis is on narrow band images derived from the NIRSpec spectroimaging datacube (see Appendix~\ref{appendix_ambient_field}). These data provide snapshots of the disc continuum emission at reference wavelengths where the ambient emission field is weak and the silhouette of the disc is seen in extinction against this continuum at nearby wavelengths. The images constrain the disc model.
\section{Modelling the Flying Saucer}
In order to better understand the NIRSpec observations and reproduce the wavelength dependence in the observed images, we describe the Flying Saucer with a radiative transfer model using the RADMC-3D\footnote{https://www.ita.uni-heidelberg.de/$\sim$dullemond/software/radmc-3d/} software \citep{Dullemond2012}. The modelling applied to this disc uses the same framework and parameter definitions  as previously described for the benchmark model of the edge-on disc Tau~042021 in \cite{Dartois2025}, and in Appendix~\ref{appendix_model_parameters}.
For the Flying Saucer model, two additional components are required: (i) an external field to simulate the extended ambient astro-PAH  emission field , (ii) a foreground extinction. 
Two icy grain populations are required: a small-grain population of sizes 0.005 to  10 $\rm\mu$m with external midplane radius $\rm r_{out}$ = 235~au; and a settled-grain population of 15 to 3000 $\rm\mu$m with external midplane radius $\rm r^{settled}_{out}$ = 190~au. This $\rm r^{settled}_{out}$ matches that reported by \cite{Guilloteau2016, Guilloteau2025}. 
Final model images are presented and interpreted following radiative transfer, ray tracing, and convolution with beam PSFs.
\section{Constraints on the disc size and dust properties}
The model of the Flying Saucer exhibits a very good overall agreement with the observational spectral imaging, 
as can be seen in Fig.\ref{fig_modele}, where we present the model convolved to the JWST spatial resolution via application of the PSF. 
The overall shapes and intensities of the lobes and the extincted disc midplane are well reproduced. The full resolution model is shown on a one colour scale in Appendix~\ref{appendix_onecolour}.
%-------------------------------------------------------------
%                        A figure as large as the column width
%-------------------------------------------------------------
\begin{figure}[ht!]
\centering
\includegraphics[width=0.9\hsize]{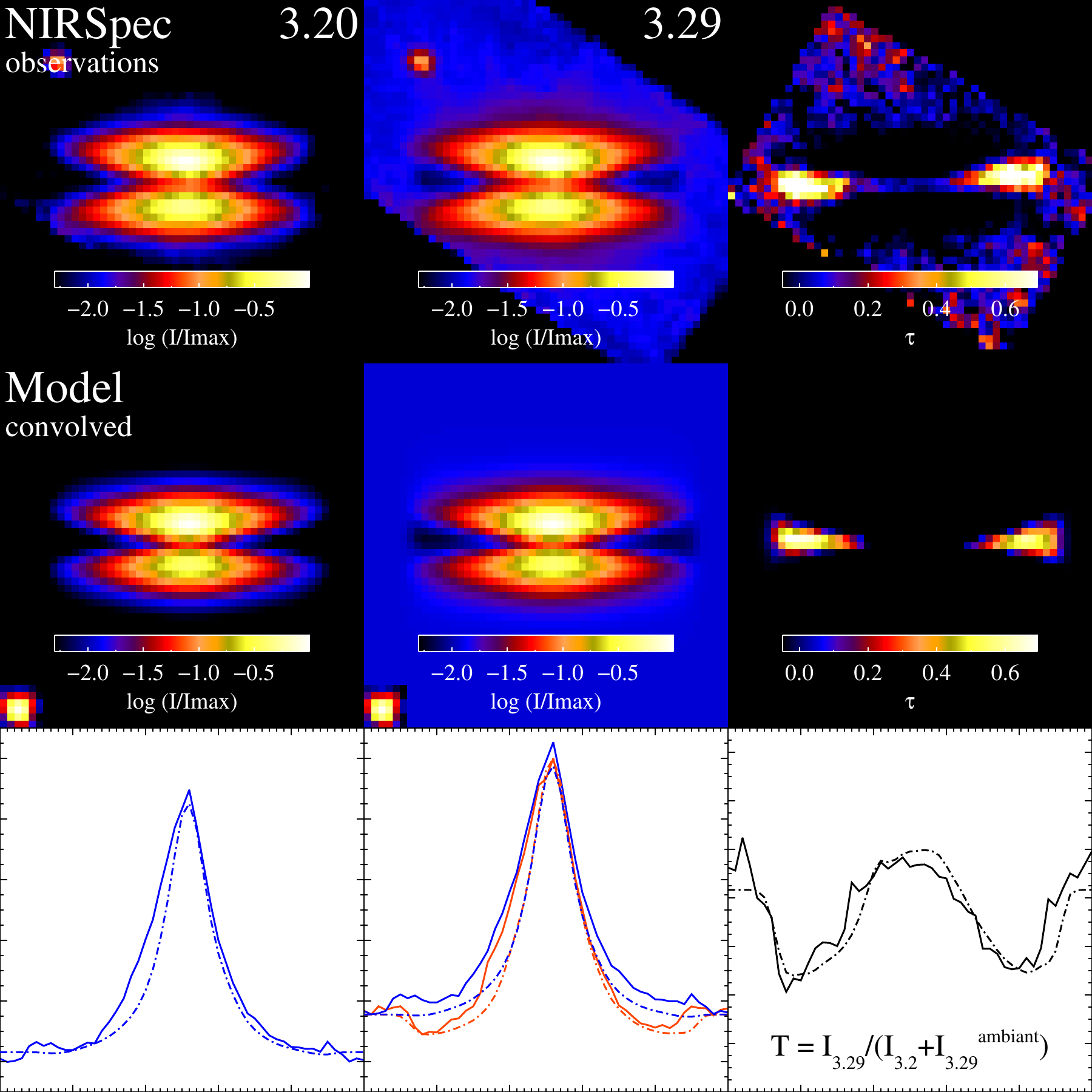}      \caption{Observations and models of disc emission/silhouette. Left and centre columns show the emission at 3.2$\mu$m ($\rm{I_{3.2}}$, no astro-PAH background) and 3.29$\mu$m ($\rm{I_{3.29}}$, peak astro-PAH background), respectively.  The right column shows the extincted optical depth $\tau$=$\rm{-ln(T)}$=$\rm{-ln(I_{3.29}/(I_{3.2}+I_{3.29}^{ambient}))}$.  
The bottom row shows profile cuts along the midplane comparing observations (solid lines) and models (dashed lines). Blue lines (left) correspond to the 3.2~$\mu$m cut. Red lines (centre) are the 3.29~$\mu$m cut. Blue cuts include an offset corresponding to the astro-PAH background contribution, i.e. the expected cut if no silhouette extinction were present. Black lines (right) show T=$\rm I_{3.29}/(I_{3.2}+I_{3.29}^{ambient}$) along the midplane cut in both model and observations.}
\label{fig_silhouette}
\vspace{-0.7cm}
\end{figure}
%-------------------------------------------------------------
The 7.7~$\mu$m MIRI archival image, included for completeness in Fig.~\ref{fig_modele}, shows much stronger background continuum emission than that at the lower wavelengths measured by NIRSpec. 
We determine spectroscopically that the emission from astronomical PAHs dominates the emission field at $\sim$~3.29~$\mu$m (Figure \ref{fig_external_field_NIRSpec}), and therefore must also contribute emission in the $\sim$8~$\mu$m region. 
In addition, it is probable that the 7.7$\mu$m emission field contains a continuum contribution from very small dust particles in the Ophiuchi cloud complex, as recorded in Spitzer IRAC band 4 images at 8~$\mu$m \citep{Evans2009}. 
In wavelength ranges covered only by photometric observations (e.g. $\sim$8~$\mu$m) the model is slightly less accurate, given that further components (i.e. atomic or molecular lines) may possibly also be contributing an unknown fraction of the flux captured by the F770W filter. Spectroscopy would be required to disentangle the various components. Nevertheless, for the ambient field adopted in the model (Fig.~\ref{fig_external_field_NIR_MIR}), we obtain agreement with the observation, including the distinct silhouette extinction pattern near the midplane. 
To precisely characterise the disc extinction, we produced high-contrast images via synthetic imaging filters. The second column in Fig.~\ref{fig_modele} is centered on the ambient emission from a narrow rovibrationally excited H$_2$ line (at 3.004~$\mu$m). Despite its low signal-to-noise ratio, the disc is clearly detected in silhouette along the midplane, with the continuum of the emission line corresponding to a maximum of a wide water ice absorption band (see Appendix~\ref{appendix_ambient_field}).
In another wavelength region of interest, one filter is  centered on the maximum of the PAH emission feature at 3.29$\mu$m (filter I$_{3.29}$, stacking images within $3.29\pm0.01\mu$m), and another centered just before the rise of the PAH stretching mode at 3.2 $\mu$m (filter I$_{3.2}$, stacking all images within $3.2\pm0.01\mu$m) (Fig.\ref{fig_external_field_NIRSpec}). The images are shown in the upper panels of Fig.\ref{fig_silhouette}. 
By dividing $\rm I_{3.29}/(I_{3.2}+I_{3.29}^{ambient})=T$, where $\rm I_{3.29}^{ambient}$ is the flux level in the surrounding ambient field from astro-PAHs, we correct for the disc scattered light and isolate the observed extinction silhouette ($\rm\tau{=-ln(T)}$), shown in the upper right panel of Fig.\ref{fig_silhouette}.  The model well reproduces the disc's vertical and horizontal profiles, both in terms of the apparent optical depth profile along the midplane and the outer radius of the silhouette disc (Fig.~\ref{fig_silhouette}).

\begin{figure}[h!]
\centering
\includegraphics[width=0.8\columnwidth]{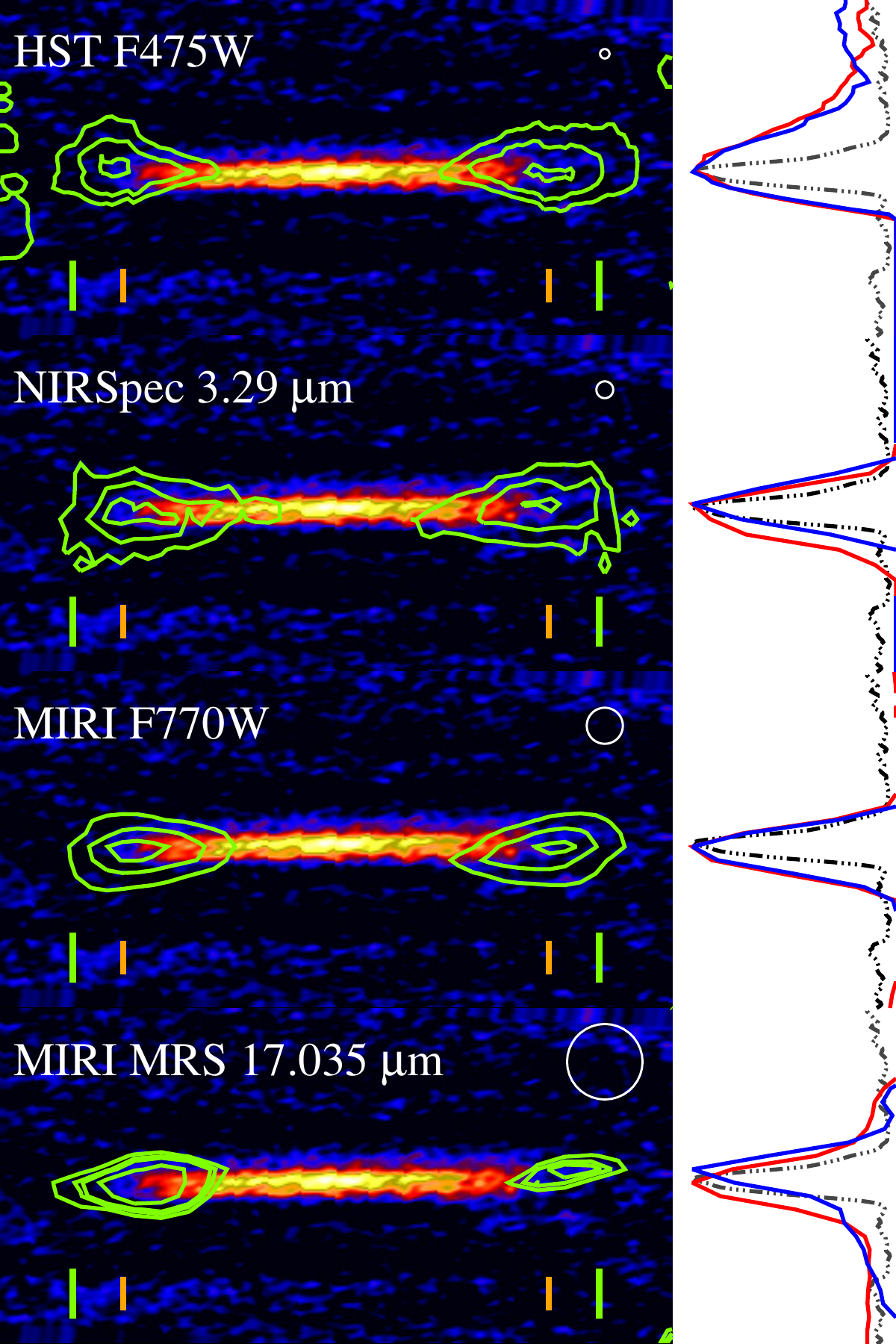}
\caption{Silhouette of the Flying Saucer disc seen against the ambient field at different wavelengths (green contours) plotted over the ALMA Band 7 millimetre dust disc image (colormap). From top to bottom, the disc silhouette is shown at 0.47 (HST F475W), 3.29 (NIRSpec), 7.7 (MIRI F770W) and 17.035 (MIRI/MRS) microns.  Image and contours are rotated by 3$\rm^o$ and the image is 5'' wide. In each panel, two sets of vertical tick marks denote the outer radii determined in the model for the small grain (235 a.u., green) and large grain (190 a.u., orange) populations, assuming a distance of 120 pc. The FWHM of the PSFs are shown as circles. At each wavelength, normalised vertical profiles are plotted to the right of the image. These are the projection across the full disc diameter for the ALMA image (dot-dashed, black) and cuts taken through the thickest point of the left lobe of the silhouette (red) and right lobe of the silhouette (blue). In both the radial and vertical directions, the disc silhouette clearly has a greater extent than the continuum image.}
\label{fig_overlap}
\vspace{-0.7cm}
\end{figure}

The detection of the disc in extinction offers an opportunity to directly probe the extent of the Flying Saucer midplane. For comparison, our observations are plotted over 
ALMA band 7 continuum data 
in Fig.~\ref{fig_overlap}. Using a combination of several ALMA data sets, a large-grain disc size of $\sim$190 au was recently measured \citep{Guilloteau2025}. 
It is clear that the disc midplane measured in extinction at 3.29~$\mu$m by NIRSpec has a larger extent compared to the millimetre disc measured by ALMA.
When comparing the NIRSpec data to archival images of the extincted disc measured at 0.47~$\mu$m (HST) and 7.7~$\mu$m (MIRI), all three also exhibit a similar spatial extent, corresponding to a radius of about $235\pm10$ au (for an adopted distance of 120 pc). 

In addition to the larger horizontal extent, it is clear that the vertical extent of the extincted disc measured by NIRSpec is also larger than that measured by ALMA.  Note that for this comparison, we must focus on the outer radial region of the disc.
The silhouette (extinction) has a dumbbell shape, since the contribution of disc scattering/emission prevents us from measuring the disc extent towards the inner radial region.  
These observations together show clearly that the grains probed across the entire visible to mid-IR range, spanning sub-micron to tens of micron sizes, are not settled to a large extent, in agreement with the full radiative transfer model we present in this work, and contrary to the larger grains probed with ALMA.
Interestingly, in our dust model the small-grain scale height at 100 au (about 7.8 au) is lower than that required to model observations of gas phase ALMA CO (2-1) transitions \citep[13.5 au at 100 au;][taking into account the $\sqrt{2}$ factor difference in parametrisation]{Guilloteau2025, Dutrey2025a}. In our model, the small-grain is tracing primarily icy dust grains. The right panel of Fig.\ref{fig_external_field_NIRSpec} shows the presence of water ice at high vertical altitude in the disc, especially in the upper lobe, thus requiring that icy grains are used up to high altitude in the radiative transfer model of this disc. This would correspond to elevations where the gas phase CO depletion begins in \citep{Guilloteau2025}, i.e. the onset of H$_2$O/CO ice (snow) surfaces. 
In contrast, some millimetre CO transitions may be probing the more extended atmosphere which deviates from a Gaussian hydrostatic profile and dominates at large scale heights. This explains why we also require the inclusion of a small fraction of the disc's small-grain dust population at very high elevations  (see Appendix \ref{appendix_model_parameters} parameters).  It should be noted that some of the parameters in such disc models could be interconnected/degenerate; for example changing the surface density exponent parameter p will influence the scale height exponent parameter h while still providing an acceptable fit.

The large-grain scale height in our model is only $\sim$a third or a quarter as large as the small-grain scale height, signifying a settling of millimetre-sized grains to the midplane.  The small-grain population responsible for most of the NIR-MIR emission component, is found to require grains up to at least 10$\mu$m in size, in agreement with \citep{Pontoppidan2007}.  
This lends further evidence that the dust grains up to tens of microns are not settled, contrary to the (sub)-millimetre grains, which helps constrain the role of turbulence and vertical mixing in discs \citep{Villenave2020,Sturm2024,Duchene2024,Dartois2025,Tazaki2025}.
%_____________________________________________________________
%                        A figure as large as the column width
%-------------------------------------------------------------
%-------------------------------------------------------------
Overall, the possibility of comparing highly spatially resolved JWST observations of dust, gas and ice to those obtained by instruments spanning wavelengths from the visible to the millimetre is currently offering a comprehensive vision of complex phenomena in discs, for example constraining the spatial extent of cavities \citep{Sturm2024} and nested small dust grain and gas outflows \citep{Duchene2024,Dartois2025}, dust settling \citep{Villenave2020} or condensation \citep{McClure2025} and tracing disc silhouettes \citep{Ballering2025}.
These new JWST observations of Flying Saucer have unambiguously revealed the small-grain disc extent in silhouette against an ambient astro-PAH background. The silhouette disc is also seen in extinction against vibrationally excited H$_2$ emission, although at a lower signal-to-noise ratio.
The detection of the disc in extinction offers a rare opportunity to directly probe the full horizontal extent of the disc, without relying on scattered light.  The silhouette disc extends to larger radii than the large grains observed by ALMA, thus providing complementary constraints on the overall disc structure and grain settling behaviour.
We observe a distinctive dumbbell shape to the silhouette disc, which reflects the vertical extent of the small-grain population. The similarity of the silhouette disc shape across visible to mid-IR wavelengths confirms that grains up to tens of microns in size are not settled, even as far out as the disc outer truncation radius. 
Follow-up analysis will be performed to retrieve the full potential of this dataset, including finer spectroscopic exploration.
\begin{acknowledgements} 
This work is based on observations made with the NASA/ESA/CSA James Webb Space Telescope and the NASA/ESA Hubble Space Telescope. The data were obtained from the Mikulski Archive for Space Telescopes at the Space Telescope Science Institute, which is operated by the Association of Universities for Research in Astronomy, Inc., under NASA contracts NAS 5–26555 and NAS 5-03127.
Support for Program number 5299 was provided by NASA through a grant from the Space Telescope Science Institute, which is operated by the Association of Universities for Research in Astronomy, Inc., under NASA contract NAS 5-03127.  This paper also makes use of data from the ALMA program 2022.1.00742.S. ALMA is a partnership of ESO (representing its member states), NSF (USA) and NINS (Japan), together with NRC (Canada), NSTC and ASIAA (Taiwan), and KASI (Republic of Korea), in cooperation with the Republic of Chile. The Joint ALMA Observatory is operated by ESO, AUI/NRAO and NAOJ.  ED and JAN acknowledge support from the Thematic Action `Physique et Chimie du Milieu Interstellaire' (PCMI) of INSU Programme National `Astro', with contributions from CNRS Physique \& CNRS Chimie, CEA, and CNES. JCS is supported by the Heising-Simons Foundation through a 51 Pegasi b Fellowship.
ZYL is supported in part by NASA 80NSSC20K0533 and NSF AST-2307199. 
LM, FMe, received funding from the European Research Council (ERC) under the European Union's Horizon Europe research and innovation program (grant agreement No. 101053020, project Dust2Planets).
Part of this research was carried out at the Jet Propulsion Laboratory, California Institute of Technology, under a contract with the National Aeronautics and Space Administration (80NM0018D0004).
\end{acknowledgements}
\bibliographystyle{aa}
\bibliography{JEDIce_for_arXiv.bib}
\begin{appendix}
%%%%%%%%%%%%%%%%%%%%%%%%%%%%%%%%%%%%%%%%%%%%%%%%%%%%%%%%%%%%%%%
% In the PDF output, floats should be placed
% under their own appendix, not before the title, nor after the
% title of the next appendix.

% In short appendices, onecolumn floats (\figure*
% or \table*) will generate a blank page.
% To prevent this behaviour, a few examples are provided here. 

% In case you have a lot of floating objects for little text and the 
% LaTeX engine moves the floats away from their context, the command
% \FloatBarrier of the “placeins” package will empty the
% float buffer and place all stored floats in the continuity.

% If you still encounter problems with wide floats placement,
% just use the onecolumn environment throughout the appendices.
%%%%%%%%%%%%%%%%%%%%%%%%%%%%%%%%%%%%%%%%%%%%%%%%%%%%%%%%%%%%%%%
%____________________________________________________________
%       Wide floats at the start of an appendix: first method
%-------------------------------------------------------------
% To prevent a blank page after the start of an appendix:
% - Switch to one \onecolumn first
% - Declare the section title
% - Declare the onecolumn float with the parameter [h!]
% - Revert to \twocolumn at the end of the section
\onecolumn
\section{Archive data}
%
%_______________________________________
%  A figure as large as the column width
%---------------------------------------
   \begin{figure*}[h!]
   \centering
   \includegraphics[width=\columnwidth, trim={0 0cm 0cm 0cm},clip]{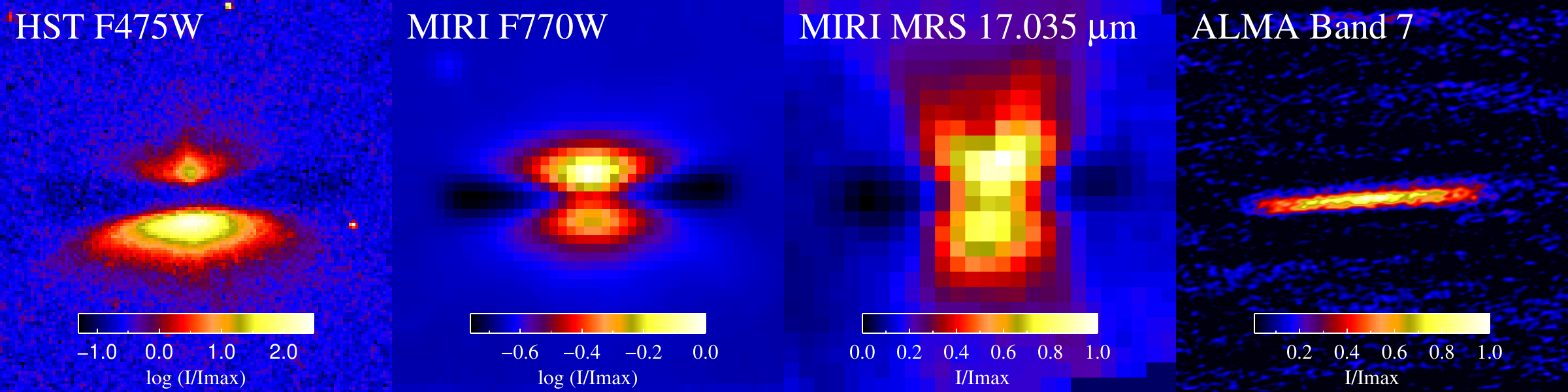}
      \caption{Archival data showing the disc in broadband imaging from HST (visible, extinction), JWST MIRI/Imager (MIR, extinction), and ALMA (Band 7, emission), and a narrowband filter derived from spectroimaging with JWST MIRI/MRS (MIR, extinction).}
         \label{fig_archive_data}
   \end{figure*}
%------------------------------------------
%
\section{Adopted external ambient field}\label{appendix_ambient_field}
\begin{figure*}[h!]
   \centering
   %trim=left lower right upper
   \includegraphics[width=0.275\hsize, trim={0cm 0cm 0cm 0},clip]{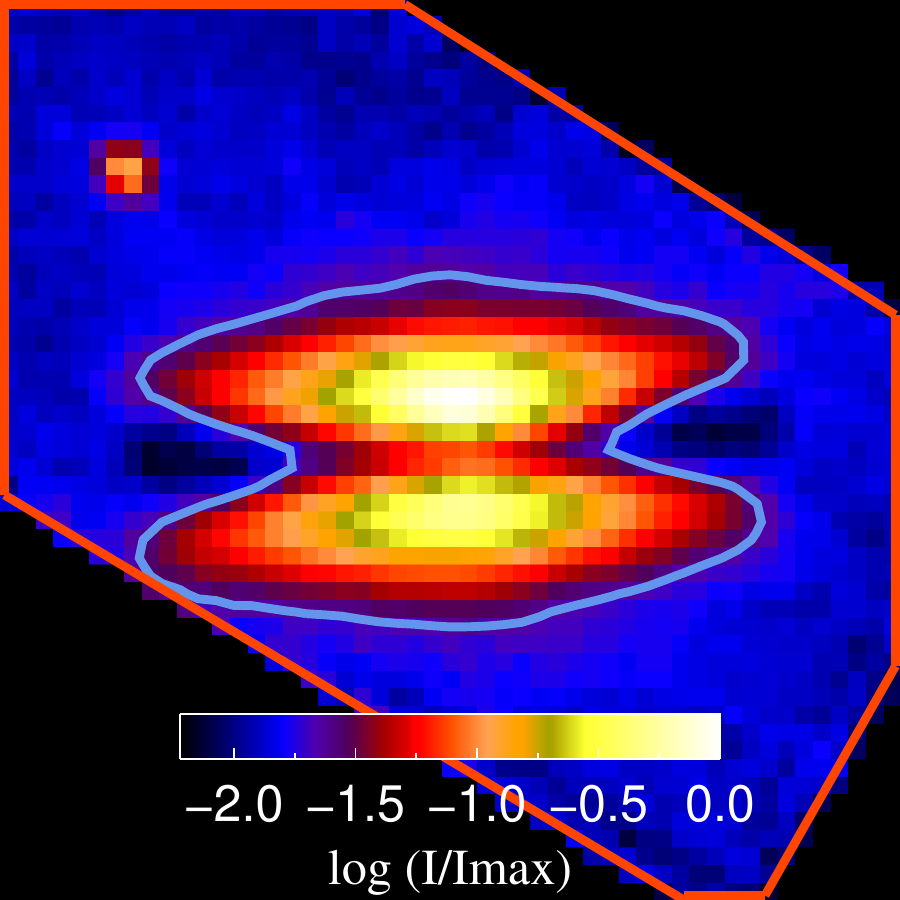}
\includegraphics[width=0.39\hsize]{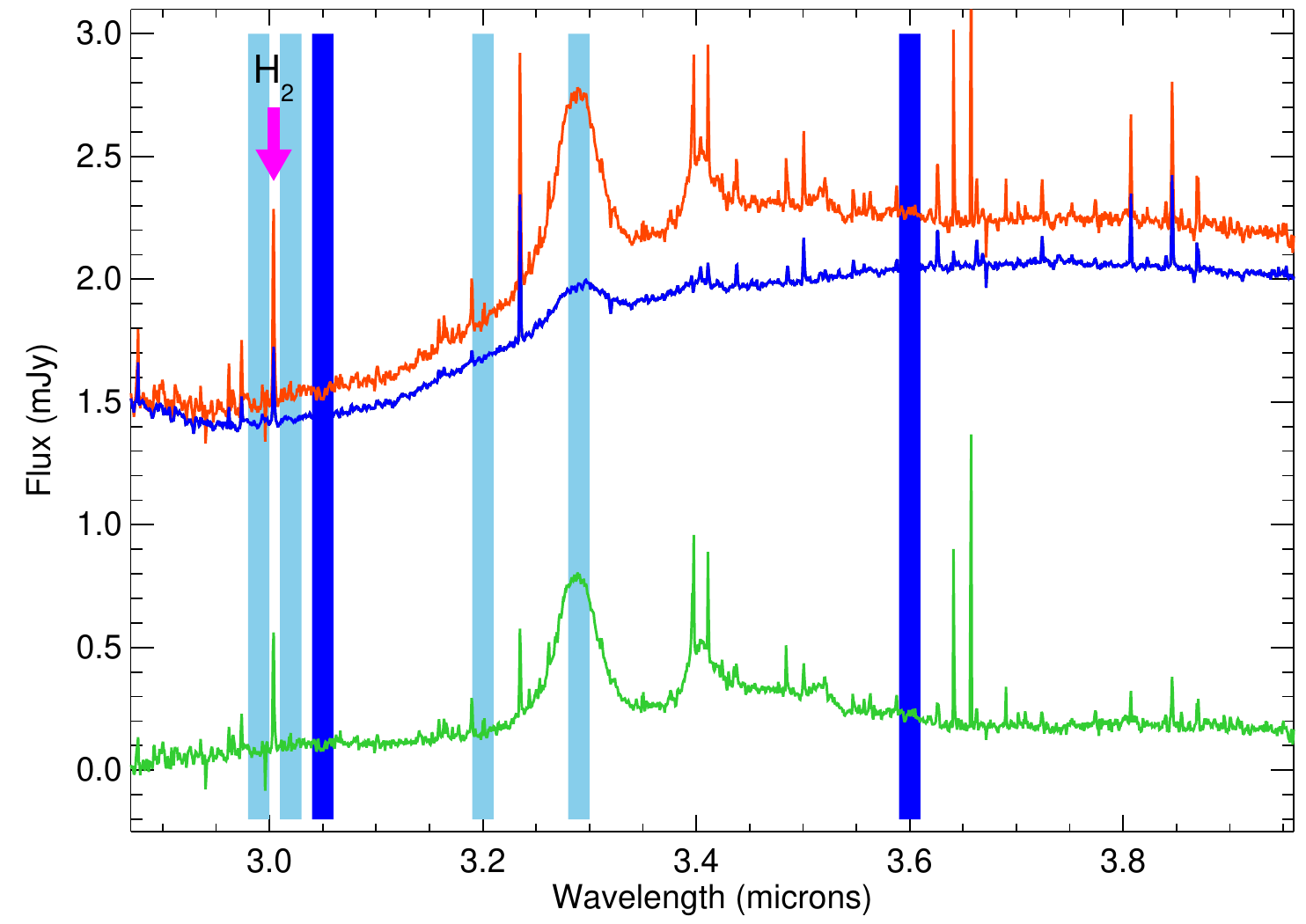}
\includegraphics[width=0.275\hsize]{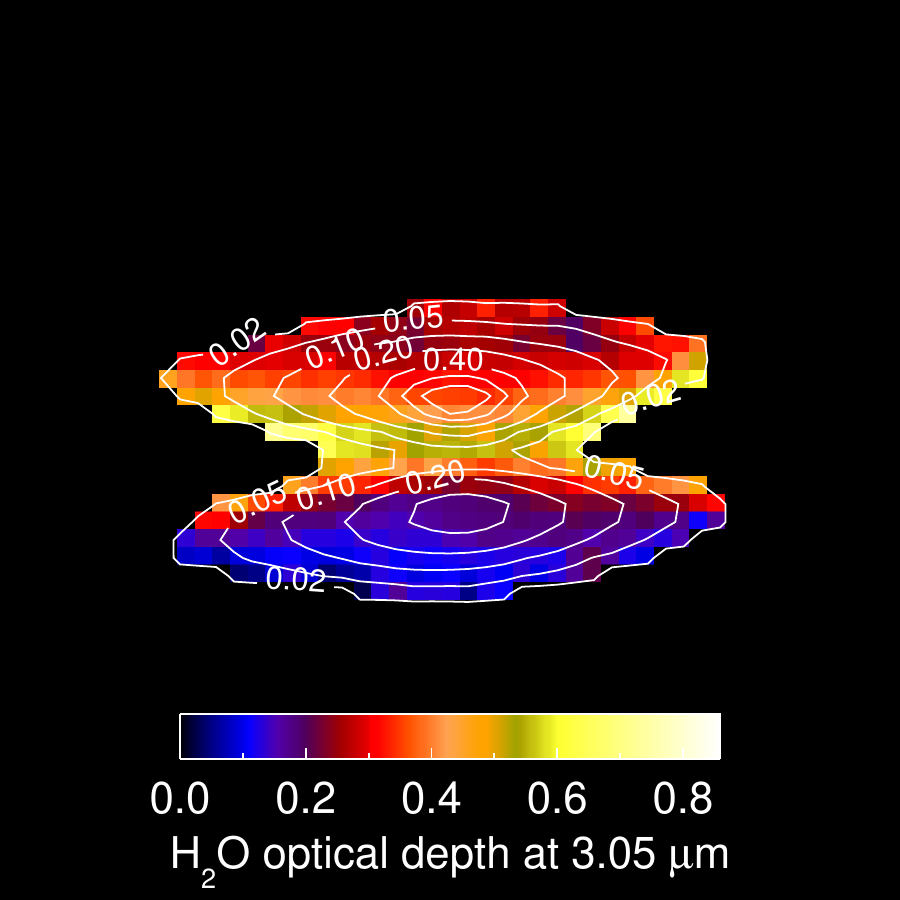}
      \caption{Left: Footprints used to extract (i) the disc emission (blue contour) and (ii) the flux in the entire image (red contour). The ambient field contribution is derived by subtracting one from the other. Middle: Spectrum of the entire field (red); spectrum within the disc footprint (blue); difference spectrum to show the ambient emission field (green). The vertical filled regions show the bandpasses of the tailored filters which where used to make the images shown in Fig.\ref{fig_modele} (light blue, at (2.99 and 3.02), 3.2 and 3.29$\mu$m). For the H$_2$ image at 3.004$\mu$m, only the three channels in which the emission line contributes were used (around the position shown with the magenta  arrow), whereas the emission line-free filter, termed `2.99~$\mu$m', makes use of the channels on both sides of the line position to estimate the disc emission around this H$_2$ line.
      Right: optical depth image ($\rm -ln(I_{3.05}/I_{3.6})$) made using the two filters shown in dark blue in the middle panel and constructed using channels  around 3.05 and 3.6 $\mu$m. The optical depth image clearly show the presence of water ice absorption (OH stretching mode absorption at 3.05~$\mu$m) at high vertical altitude in the disc, especially in the upper lobe, requiring icy grains to model the disc radiative transfer.}
    \label{fig_external_field_NIRSpec}
   \end{figure*}

The illuminating external ambient field in our model in the NIRSpec wavelength range is derived directly from JWST observations well away from the disc. Physically, this field is the result of emission from the astro-PAH component in the nearby cloud. Details of the footprint used are shown in Fig.\ref{fig_external_field_NIRSpec}(left). This figure also show that the CH stretching mode emission observed in the image is not arising from the disc itself by comparing the total flux in the image to a footprint enclosing the disc emission (middle). 
The right panel of Fig.\ref{fig_external_field_NIRSpec} displays the apparent optical depth observed by comparing the images at the maximum extinction level in the water ice OH stretching mode band at 3.05~$\mu$m to a reference channel in the adjacent continuum at 3.6~$\mu$m (filters shown in dark blue in the middle panel of Fig.\ref{fig_external_field_NIRSpec} and constructed using channels around 3.05 and 3.6 ~$\mu$m), by forming the optical depth image ($\rm -ln(I_{3.05}/I_{3.6})$).
The optical depth image clearly show the presence of water ice absorption 
at high vertical altitude in the disc, especially in the upper lobe, requiring icy grains to model the disc radiative transfer.
%_____________________________________________________________
%                        A figure as large as the column width
%-------------------------------------------------------------
%-------------------------------------------------------------

\begin{figure}[h!]
   \centering
      \includegraphics[width=0.5\hsize]{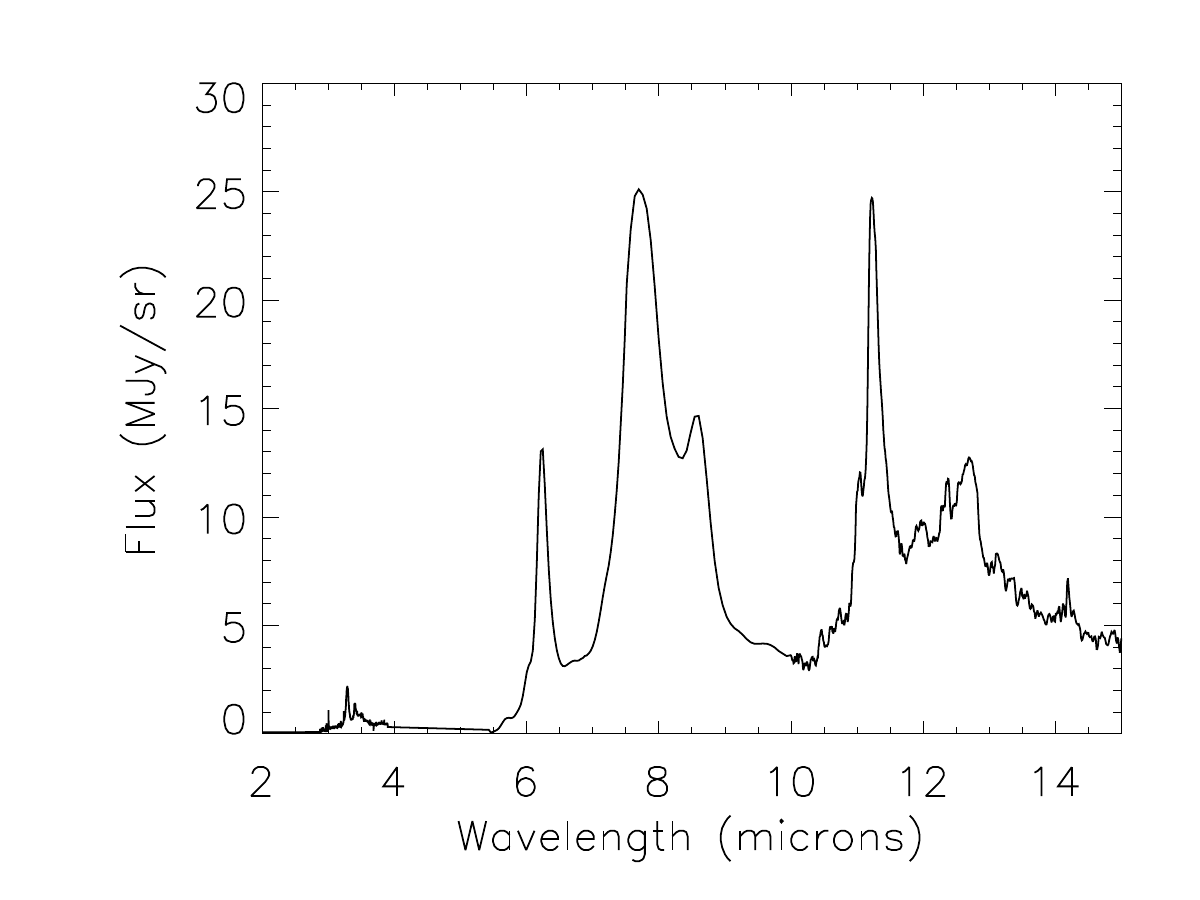}
      \caption{Ambient field adopted in the modelling. The NIR (2.9--4 $\mu m$) region was directly obtained from NIRSpec observations in this work (see Fig.\ref{fig_external_field_NIRSpec}). The mid-IR part (5.5-15$\mu$m) is a Spitzer spectrum, measured around the disc HD97300, whose flux was scaled to mimick the expected ISRF flux in the 7.7$\mu m$ JWST broadband filter. In the mid-IR, the contribution to an underlying continuum from the emission of very small grains \citep[e.g.][]{Compiegne2011} would be required to derive a more accurate ambient field emission profile.}
         \label{fig_external_field_NIR_MIR}
   \end{figure}

In the Mid-IR range, in the absence of more spectral information, we scaled the expected contribution by astronomical PAHs to reach the $\sim$12MJy/sr intensity in the continuum in the F770W MIRI filter, as shown in Fig.\ref{fig_external_field_NIR_MIR} . A more detailed ambient field would require spectral information to be able to measure the continuum contribution from very small grains on top of which the contribution of the astronomical PAH mid-IR bands would lie. 

%_____________________________________________________________
%                        A figure as large as the column width
%-------------------------------------------------------------
%-------------------------------------------------------------
\section{Benchmark disc model}
\label{appendix_model_parameters}
%_____________________________________________________________
\begin{figure}[!htb]  
\center
\includegraphics[width=0.5\hsize]{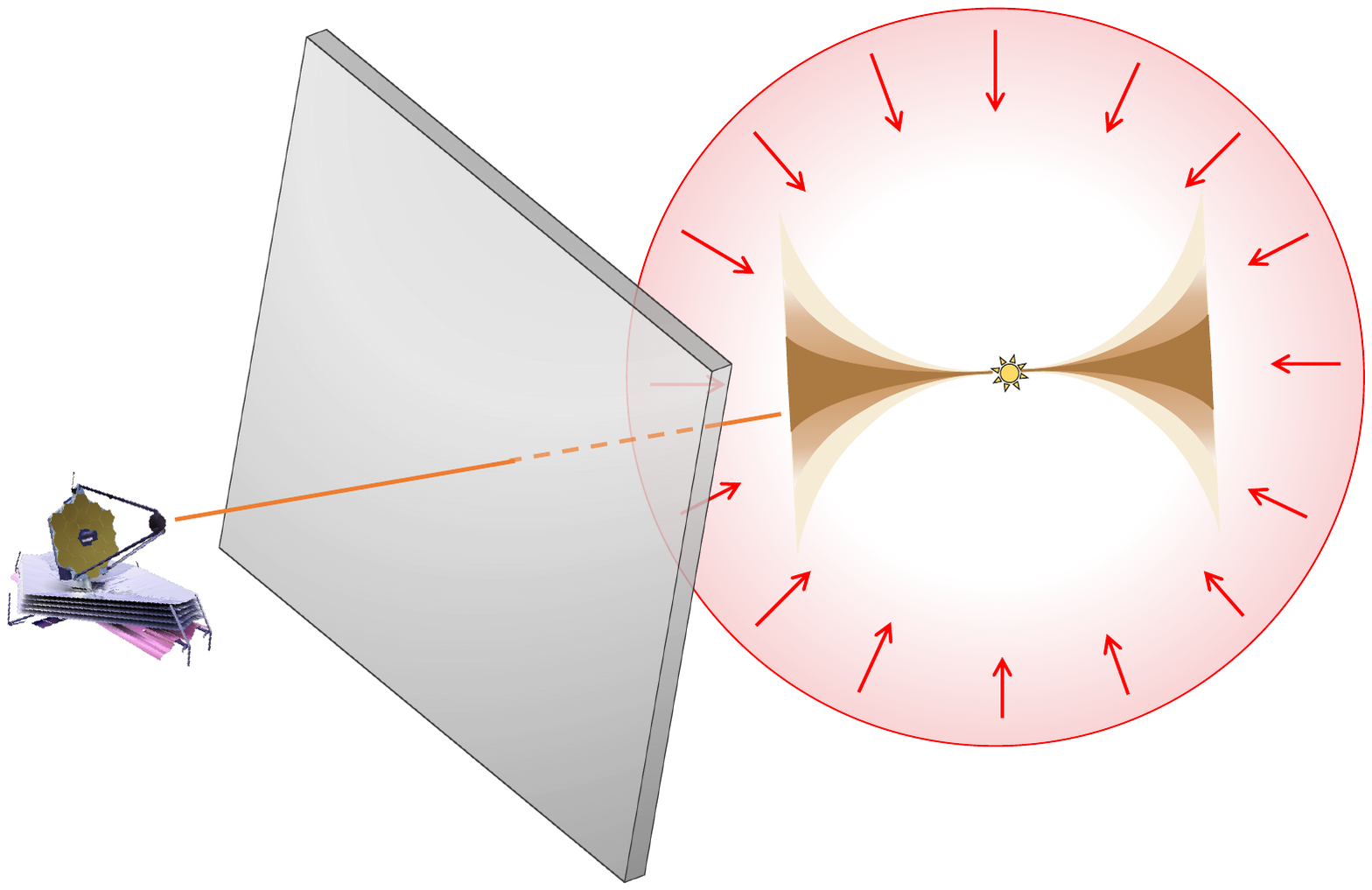}
\caption{Schematic view of the Flying Saucer radiative transfer calculation. The disc is immersed in an external astro-PAH ambient emission field modelled as a spheroidal envelope, illustrated in red, and lies behind a foreground extinction, illustrated in grey.}
\label{figure_schematique}
\end{figure}
%-------------------------------------------------------------
The best-fit benchmark model we obtained is shown in Fig.\ref{fig_modele}. This model reproduces the spatial distribution in emission and extinction from the NIR to the millimetre. The definition of all model parameters and their values in the benchmark model are given here. %in Appendix~\ref{appendix_model_parameters}. 
A schematic view of the different model components used to model the Flying Saucer is shown in Fig.\ref{figure_schematique}. For a close-up of the disc grain spatial distribution scheme, we refer to the benchmark model of Fig.5 in \cite{Dartois2025}. 
The external field spectrum was estimated in the NIRSpec range by summing sky spaxels in the external continuum around the Flying Saucer, and is shown in Fig.\ref{fig_external_field_NIRSpec}. It was extended into the mid-IR (Fig.\ref{fig_external_field_NIR_MIR}) using the astronomical PAH emission band spectrum of the HD97300 disc retrieved from the Spitzer archive (PID 2; Astronomical Observation Request 12697088, Low and High resolution). This Mid-IR spectrum was scaled to achieve an integrated flux compatible with (i) the MIRI F770W region outside of the disc, and (ii) the archival IRAC 8$\mu$m filter extended maps of the Spitzer legacy program c2d \cite{Evans2009}. In addition, as discussed previously in the pioneering work on this source from \cite{Grosso2003}, a foreground interstellar dust extinction must be taken into account. For this extinction contribution, we added an $\rm R_V=5.5$ Milky Way extinction curve taken from \cite{Draine2003}, adjusted to the expected visual extinction of around A$_V=$ 2.

In order to analyse the observations, we use a benchmark disc model with several layers of structural complexity as described below.
The classical, so-called `standard' disc model making the core of the model is based on an axisymmetric flaring disc around a young star, described using classical parameters.
The surface density of the disc is parametrised with
\begin{equation}
\rm    
\Sigma(r) = \Sigma_0 
\left(\frac{r}{r_{\text{0}}}\right)^{-p} 
\exp\left(-\left(\frac{r}{r_t}\right)^{2-p}\right)
\end{equation}
where r is the radial distance to the star, $r_{\text{0}}$ is a reference radius and $\rm \Sigma_0$ the surface density normalisation at this reference radius. p is the power law exponent describing the radial variation, positive if the surface density decreases with radius. The density is tapered by an exponential edge with reference radius $r_{\text{t}}$, following the viscous disc model theoretical solution (Lynden-Bell \& Pringle 1974).

The vertical density distribution, under hydrostatic equilibrium, is given by
\begin{equation}
\rm        \rho(r, z) = 
\frac{\Sigma(r)}{H(r)\sqrt{2\pi}}
\exp\left(- z^2 / 2 H^2(r)\right)
\label{eqn_vertical_density_hydro}
\end{equation}
with the density at radial distance $\rm r$ and vertical distance $\rm z$ from the midplane,
and where $\rm H(r)$ is the vertical scale height under a vertical isothermal hypothesis, whose radial variation is given by
% isothermal
\begin{equation}
H(r) = H_0 \left(\frac{r}{r_0}\right)^{h}
\end{equation}
where $\rm H_0$ is the scale height at the reference radius $\rm r_0$. h is the power law exponent describing the scale height radial variation, $>1$ for flaring discs.
The midplane density radial variation evolves thus with a power law as $\rm s = -(p + h)$.\\

The disk extends from its inner radius $\rm r_{in}$, defined by the sublimation temperature of the refractory components to the outer radius $\rm r_{out}$. 
The large grain distribution possesses its own (lower) scale height $\rm H^L_0$.\\

To model the effect of an extended atmosphere, i.e. an atmosphere deviating from the pure hydrostatic equilibrium, we adopt a simple configuration by slightly modifying the vertical density profile in a way which allows empirically fitting some of the protoplanetary disc expected prescriptions models, some cited in \cite{Dartois2025}. 

To do so, above a given altitude, the extended atmosphere takes precedence over the classical hydrostatic decay, and is defined by parameters $\rm \epsilon$ and $\eta$. We thus modify equation~\ref{eqn_vertical_density_hydro} by replacing it with the following extended atmosphere equation:

\begin{equation}
\rm        \rho^{e.a.}(r, z\ge z_{th}) =   
\frac{\Sigma(r)}{H(r)\sqrt{2\pi}} \;\;
exp(-\epsilon) \exp\left(-\alpha\right) \;;\; \alpha = \eta\frac{|z|}{\sqrt{2}H(r)}
\label{eqn_vertical_density_modified}
\end{equation}
The transition where equation \ref{eqn_vertical_density_hydro} equals equation \ref{eqn_vertical_density_modified} occurs at 
\begin{equation}
\frac{z_{th}}{H(r)} = \frac{ \eta + \sqrt{4 \epsilon+\eta^2}}{\sqrt{2}}
\end{equation}

To obtain the benchmark model fit shown in this study, we used a $\rm\chi^2$ minimisation scheme simultaneously minimising the sum of several reduced $\rm\chi^2$s, taking into account the following major constraints: the log intensity of the observed image at 3.29 $\mu$m, the silhouette cuts along the midplane at 3.29 and 7.7 $\mu$m, the cut along the midplane after dividing the 3.29 $\mu$m image by 3.2 $\mu$m image, as shown in Fig.\ref{fig_silhouette}, and the fluxes of the different images. This minimisation scheme provides a  balance of all these different observational constraints. The minimisation was performed using a Nelder-Mead downhill simplex method algorithm, starting from various initial conditions within the parameter space provided to the algorithm to verify that the final result is not a local minimum. The parameter space exploration was made over the following ranges: inclination = $\rm 86^{\circ}-89^{\circ}$, hydrogen number density at the reference radius $r_0$ of 100 au, $n_{\text{H}_{2}}(100~\text{au}) = 10^7-5\times10^8$~cm$^{-3}$, $\rm p = 0.0-1.0$, $\rm r_{in}$ = 0.07-10~au, $\rm r_{out}$ = 200-250~au, $\rm H_{0} = 4-13$~au, $\rm h = 1.0-1.5$, $\rm H_{0}^L = H_{0}/1-5$.
The grain size distribution boundaries (a$\rm_{min}$, a$\rm_{max}$, a$\rm^{settled}_{min}$, a$\rm^{settled}_{max}$) were explored on a discrete grid: 0.005, 0.1, 0.3, 0.7, 1, 2, 3, 5, 10, 15, 20, 30, 40, 50, 60, 100, 1000, 3000 $\mu$m.

The main parameters of the benchmark RADMC3D model shown in the article are: T$_{star}= 3700$~K, a disc inclination of 87.8$^{\circ}$, the hydrogen number density, at the reference radius $r_0$ of 100 au, $n_{\text{H}_{2}}(100~\text{au}) = 1.05\times10^8$~cm$^{-3}$, $\rm p = 0.23$, $\rm r_{in}$ = 0.07~au, $\rm r_{out}$ = 235~au, $\rm H_{0} = 7.8$~au, $\rm h = 1.23$, $\rm H_{0}^L = 1.8$~au. 
The NIRSpec instrumental PSF is on the order of 0.1$^{\prime\prime}$ around 3$\mu$m, corresponding to  the $\pm\sim$10 au reported in the main text. 
The model includes the possibility to include a lower density cavity $\rm r_{cav}$. 
The model result with $\rm r_{cav}=1.05$ au agrees well with the recent study probing large grains with ALMA \citep[<2~au,][]{Guilloteau2025}.

Among the salient parameters of this model is the need for two grain size distributions: a small-grain population ranging in size from a$\rm_{min}$ =0.005 $\rm\mu$m to a$\rm_{max}$ = 10 $\rm\mu$m and extending to an external midplane radius of $\rm r_{out}$ = 235~au; and a settled-grain population ranging from  a$\rm^{settled}_{min}$ = 15 $\rm\mu$m to a$\rm^{settled}_{max}$ = 3000 $\rm\mu$m and extending to an external midplane radius of $\rm r^{settled}_{out}$ = 190~au. 
The ice used for the grain mantles is a H$_2$O:CO$_2$:CO:NH$_3$ mixture; optical constants are given in \cite{Dartois2022}. 
The settled large grain scale height in this model ($\rm H_{0}^{settled}$ = $\rm H_{0}/4.5$) corresponding to the ALMA Band 7 image. 
Such a settling factor can be considered as `moderate' when compared to other evolved discs \citep[e.g.][]{Villenave2020}, most probably because in the Flying Saucer disc, as suggested by \cite{Guilloteau2025}, the host star is still relatively young, which explains their finding of an $\rm H_{0}\sim6$ au in the millimetre, based on a more complete millimetre observations coverage, and considering an exponent of less than unity which implies some tapering in the outer disc.
The extended atmosphere, which extends the small grains distribution's vertical presence above the classical hydrostatic density decay, is calculated with set parameters $\epsilon=2.5$ and $\eta=1.0$. 
\section{Models on a one colour scale}\label{appendix_onecolour}
\begin{figure*}[h!]
   \centering
    \includegraphics[width=\hsize,trim={0cm 0cm 0cm 0cm},clip]{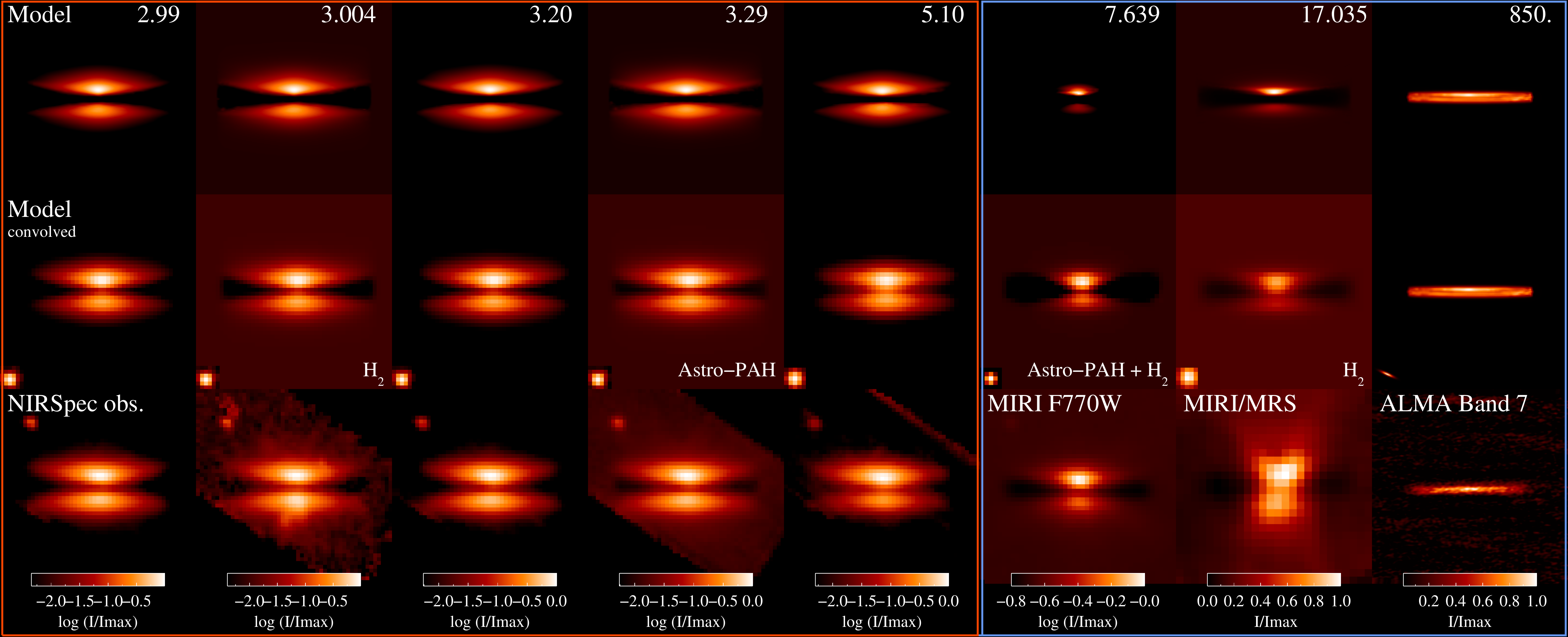}
      \caption{Benchmark model of the Flying Saucer from near-IR to mm, compared to observations on a $\rm 5^{\prime\prime} \times5^{\prime\prime}$ field of view. The central and lower rows present the same data as in Fig.\ref{fig_modele}, but on a single color scale. The upper row presents the non-convolved, full resolution model images.}
         \label{fig_modele_one_color_scale}
   \end{figure*}

This appendix contains the full benchmark model including the model images presented at full calculated resolution i.e. prior to being convolved by the instrumental Point Spread Function at the corresponding wavelengths (Fig.\ref{fig_modele_one_color_scale}, upper row).
As discussed in many articles such as \cite{Crameri2020}, 
the use of appropriate colour scales and the addition of a colorbar is mandatory to be able to interpret what is modelled without distorting the perception of the image, as well as to allow colour visually deficient people, amounting to a significant fraction of the population, to perceive the scales in the same way. To this end, we provide the models coded with two colour palettes, providing the one colour scale version in this appendix. Both colour palettes chosen should be close to being perceptually uniform. In the same effort, we systematically added colorbars for the reader to refer to.
%_____________________________________________________________
%                        A figure as large as the column width
%-------------------------------------------------------------
%-------------------------------------------------------------

%
\section{Effect of resolution on apparent optical depth in the dark lane}\label{appendix_optical_depth}

In Figure \ref{fig_silhouette} we show horizontal cuts through the observations and PSF-convolved models that compare, at the same spatial resolution, the apparent optical depth along the midplane of the disk. This appears to be concentrated/located in two lobes near the outer disc whereas at small stellocentric radius, the apparent optical depth increases because of beam dilution which merges scattered light from the disk lower local scale height.
Here, we present a comparison with the full resolution model cut, to illustrate that higher angular resolution observations would reveal a consistently deeper apparent optical depth across the midplane close to the star, whereas it asymptotically converges toward the observed depth at large stellocentric radius.

%-------------------------------------------------------------
\begin{figure}[h!]
\center
\includegraphics[width=0.25\hsize]{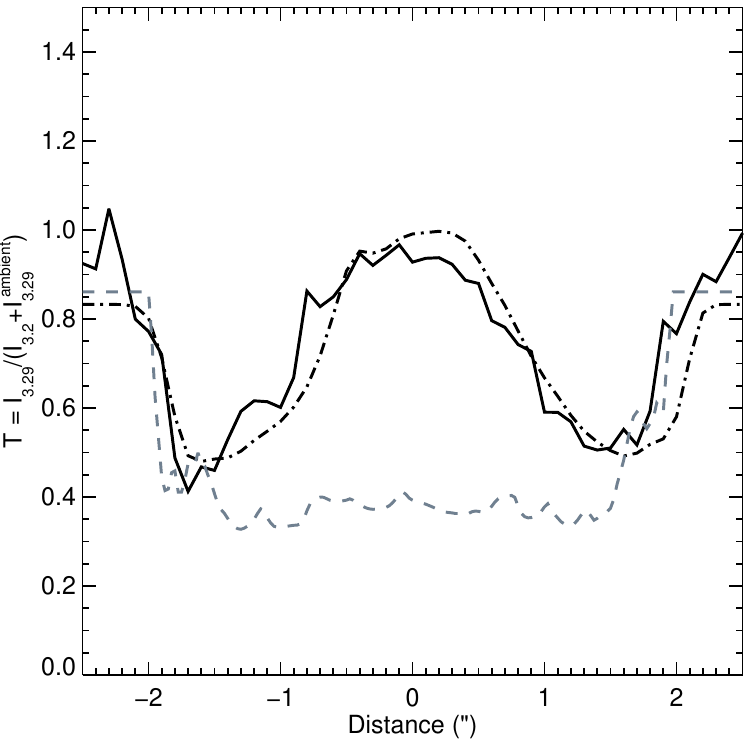}
\caption{Midplane cut for $\rm T = I_{3.29}/(I_{3.2}+I_{3.29}^{\;ambient})$, as shown in the lower right panel of Fig.\ref{fig_silhouette} (observed: solid black line; modelled: dash dot line), compared to the cut retrieved from the model images as calculated at full spatial resolution (dash line).}
\label{figure_cut_full_res}
\end{figure}
%-------------------------------------------------------------
\end{appendix}

\end{document}